\documentclass[journal, twoside]{IEEEtran}
\makeatletter

\usepackage{blindtext}
\usepackage[T1]{fontenc}
\usepackage{textcomp}
\usepackage{amsmath}
\usepackage{amssymb}
\usepackage{bm}
\usepackage{mdwmath}
\usepackage{cases}
\usepackage{graphicx}
\graphicspath{{Figure/PDF/}{Biography/PDF/}}
\usepackage[dvipsnames]{xcolor}
\definecolor{myblue}{rgb}{0,0.4980,1} 
\definecolor{myred}{rgb}{0.8706,0.1608,0.0627} 
\usepackage[caption=false,font=footnotesize,subrefformat=parens]{subfig}
\usepackage[nosort,noadjust]{cite}
\usepackage{url}
\usepackage{tabularray}
\UseTblrLibrary{counter}
\UseTblrLibrary{diagbox}

\usepackage{hyperref}
\newcommand{\colorhypersetup}{\@ifpackageloaded{hyperref}{\hypersetup{%
	bookmarksopen=true,%
	bookmarksnumbered=true,%
	pdfpagemode={UseOutlines},
	pdfstartview={FitH},%
	colorlinks=true,%
	linkcolor={myred},%
	citecolor={orange}
}}{\empty}}
\newcommand{\blackhypersetup}{\@ifpackageloaded{hyperref}{\hypersetup{%
	bookmarksopen=true,%
	bookmarksnumbered=true,%
	pdfpagemode={UseOutlines},
	pdfstartview={FitH},%
	colorlinks=true,%
	allcolors={black}
}}{\empty}}
\colorhypersetup

\usepackage{mfirstuc-english}
\MFUhyphentrue
\usepackage[version3,description,IEEEtran]{eacro}
\acsetup{%
    pages/display=all,%
    pages/seq/use=false,%
    pages/name=true,%
    pages/fill={\quad},%
    make-links=true%
}

\usepackage{algpseudocode}
\newcounter{MYalgorithmic}

{%
	\vspace*{3pt}
	\hrule height 1pt}
\newcounter{MYitem}[MYalgorithmic]

\newcommand{\MYlabel}[1]{\def\@currentlabel{\theALG@line}\label{#1}}

\newcommand{\upperroman}[1]{\uppercase\expandafter{\romannumeral#1}}

\newcommand{\myunit}[1]{%
	\ifmmode
		\mathrm{#1}
	\else
		$ \mathrm{#1} $
	\fi}
\newcommand{\murm}{%
	\ifmmode
		\text{\textmu}
	\else
		\textmu
	\fi}

\newlength{\mysinglefigwidth}
\newlength{\mymultifigwidth}
\ifCLASSOPTIONonecolumn
	\AtBeginDocument{\setlength{\mysinglefigwidth}{0.8\linewidth}}
\else
	\AtBeginDocument{\setlength{\mysinglefigwidth}{\linewidth}}
\fi
\ifCLASSOPTIONonecolumn
	\AtBeginDocument{\setlength{\mymultifigwidth}{0.5\linewidth}}
\else
	\AtBeginDocument{\setlength{\mymultifigwidth}{\linewidth}}
\fi

\makeatother
\begin{document}
\title{Digital Retina for IoV Towards 6G: Architecture, Opportunities, and Challenges}
\author{Kan~Zheng,~\IEEEmembership{Fellow,~IEEE},  Jie~Mei,~\IEEEmembership{Member,~IEEE}, Haojun~Yang,~\IEEEmembership{Member,~IEEE}, Lu~Hou,~\IEEEmembership{Member,~IEEE}, and~Siwei~Ma,~\IEEEmembership{Fellow,~IEEE}
}

\ifCLASSOPTIONonecolumn
	\typeout{The onecolumn mode.}
\else
	\typeout{The twocolumn mode.}
	\markboth{}{Author \MakeLowercase{\textit{et al.}}: Title}
\fi

\maketitle

\ifCLASSOPTIONonecolumn
	\typeout{The onecolumn mode.}
	\vspace*{-50pt}
\else
	\typeout{The twocolumn mode.}
\fi
\begin{abstract}
Vehicles are no longer isolated entities in traffic environments, thanks to the development of IoV powered by 5G networks and their evolution into 6G. However, it is not enough for vehicles in a highly dynamic and complex traffic environment to make reliable and efficient decisions. As a result, this paper proposes a cloud-edge-end computing system with multi-streams for IoV, referred to as Vehicular Digital Retina (VDR). Local computing and edge computing are effectively integrated in the VDR system through the aid of vehicle-to-everything (V2X) networks, resulting in a heterogeneous computing environment that improves vehicles' perception and decision-making abilities with collaborative strategies. Once the system framework is presented, various important functions in the VDR system are explained in detail, including V2X-aided collaborative perception, V2X-aided stream sharing for collaborative learning, and V2X-aided secured collaboration. All of them enable the development of efficient mechanisms of data sharing and information interaction with high security for collaborative intelligent driving. We also present a case study with simulation results to demonstrate the effectiveness of the proposed VDR system. 
\end{abstract}

\ifCLASSOPTIONonecolumn
	\typeout{The onecolumn mode.}
	\vspace*{-10pt}
\else
	\typeout{The twocolumn mode.}
\fi
\begin{IEEEkeywords}
Internet-of-vehicle (IoV), Sixth-generation (6G), Machine Learning (ML), and Edge Computing.
\end{IEEEkeywords}

\IEEEpeerreviewmaketitle



\section{Introduction}
\label{sec:Introduction}

\acresetall

\IEEEPARstart{T}{he} increasing popularity of autonomous driving has recently piqued the interest of both academia and industry. In particular, autonomous driving can be seen as a substantial collaboration of sequential processing, particularly sensing, perception, and decision-making. Autonomous vehicles may have a precise understanding of surrounding road conditions thanks to sensors, including Light Detection and Ranging (LiDAR), video cameras, radar, and others. The rate of perception data generated by these sensors can reach up to 2 Gbps, and for safe autonomous driving, sensory data must be processed accurately and within tight latency requirements in milliseconds. However, the needs of in-vehicle real-time processing of perception data are approaching or even exceeding the computational power of a single vehicle. Furthermore, deep machine learning methods for decision-making are computationally intensive, making real-time execution by a vehicle itself challenging. Additionally, a vehicle might be in trouble when it encounters occlusion (blind spots) and anomalous sensory data when driving. The existing paradigm of autonomous driving is mainly focused on the perception and decision-making by a single vehicle, making it difficult to overcome the aforementioned challenges~\cite{Han2023}.

Consequently, it is time to advocate for a new paradigm shift in the design philosophy of autonomous driving systems. With the rise of the Internet-of-Vehicle (IoV) towards 6G~\cite{6g-iov}, the use of Intelligent Connected Vehicles (ICVs) is emerging as a potential solution, where ICVs communicate with vehicles, roadside equipment, and cloud/edge servers to perform collaborative perception and decision-making~\cite{Nguyen2022}. This can not only broaden the perception range of the driving environment, but also fully utilize the computing resources of edge and cloud servers to obtain safer and more efficient autonomous driving decisions~\cite{Malik2021}. As a result, a systematic design should be carried out in order to employ all available resources, including those for communication and computation.
 
Most of the existing studies on collaborative perception and decision-making schemes in IoV can be roughly divided into two main categories, i.e., computing-related and communication-related ones. The former primarily focuses on understanding the environment and making correct decisions and plans, including perception, motion behavior decisions, and route planning. The latter, on the other hand, aims to efficiently exploit communication resources to let vehicles, road facilities, and users cooperate together. These two categories, however, are interrelated yet have not been jointly designed for autonomous driving in IoV.

On the other hand, the Digital Retina (DR) system was proposed to efficiently utilize communication and computing resources to process visual data in a smart city, using an edge and cloud computing architecture with multi-stream parallelism~\cite{Gao2021}. The devices of the DR system have efficient video coding capabilities to extract feature data before aggregating it to the edge and cloud in real-time. Then, the edge or cloud servers can quickly process the video on demand for tasks such as data retrieval, analysis, and mining based on the feature data. Since the DR system is mostly used for video surveillance, it did not take into account the effects of factors such as device mobility and radio resource constraints on performance. However, these are precisely the issues that must be addressed in vehicular networks. Moreover, the design for streams in the DR system did not account for any requirement for collaboration.

Therefore, motivated by the layered computing architecture of the DR system, this paper further proposes a new multi-layer and multi-stream \textit{Vehicular Digital Retina (VDR)} system while taking the characteristics of vehicular networks into account. The objective of the VDR system design is to jointly utilize various system capabilities, including perception, computing, and communication, for achieving the goals of collaborative driving in IoV. In the VDR system, vehicle collaboration is achieved by sharing the extracted knowledge rather than directly exchanging the sensory data. Through the collaborative perception of multiple vehicles, the surrounding environment can be sensed beyond the Line-of-Sight (LoS) for a more accurate understanding of the road environment. This enhanced perception enables vehicles to make safe and reliable driving decisions with the aid of vehicle-to-everything (V2X) communications. The main contributions of this paper are summarized as follows:
\begin{enumerate}
\item There are three layers in a VDR system, i.e., cloud, edge, and end layers, all of which are connected with communication links. Different from the traditional DR system, three new kinds of streams are defined to exchange different types of characterized data to facilitate collaborative learning in VDR.

\item The collaborative perception with the aid of V2X communication is then presented to enhance the perception performance of vehicles under a complicated environment. Various purposes of multi-device collaborative perception are presented, along with brief descriptions of how to achieve them.

\item The strategies of stream sharing for collaborative learning are proposed using V2X communications to efficiently utilize distributed sensory data as well as heterogeneous computing resources. In particular, multiple types of knowledge streams are identified, and they are to be shared via various transmission techniques. Different approaches for model aggregation and distribution are provided for model sharing.

\item We also present a variety of V2X-aided secured collaboration schemes to ensure the safe exchange of different streams in the VDR system. A reliable identity authentication mechanism is proposed to ensure secure and trustworthy participation in collaborative driving. Various approaches for data encryption and content security are advised for different kinds of streams based on their properties.
\end{enumerate}

\section{Framework of Vehicular Digital Retina (VDR)}

\begin{figure*}[!t]
	\centering
	\includegraphics[width=0.8\linewidth]{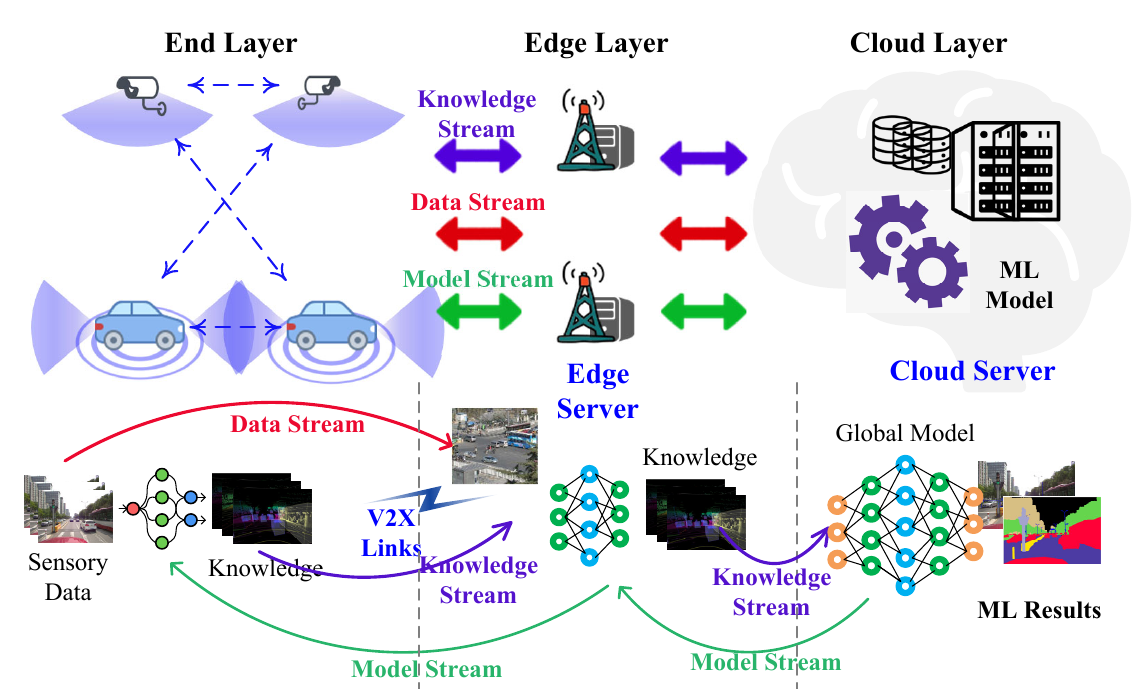}
	\caption{Illustration of the VDR System.}
	\label{fig_system}
\end{figure*}

As shown in Fig.~\ref{fig_system}, perception units at vehicles and roadside equipment have the capacity to gather environmental information. Subsequently, sensory data can undergo local processing before being efficiently and timely transmitted to edge/cloud servers. Collaboration within VDR occurs both across different layers and within individual layers. Throughout this collaborative process, information sharing and interaction are facilitated by various streams transmitted via V2X links.

\subsection{Multi-Layer Collaboration in VDR}

In general, a typical VDR system has three layers, i.e.,

\subsubsection{\textbf{End Layer}}

There are usually two types of devices in the VDR system, i.e.,
\begin{itemize}
\item \textbf{Vehicle:}  Each vehicle, such as an ICV, is equipped with a range of sensors, including cameras, Lidar, and various motion sensors. It possesses fundamental intelligent computing capabilities for extracting the knowledge of sensory data, which is subsequently subjected to further processing.

\item \textbf{Roadside Perception Equipment (RPE):}  It has roadside surveillance cameras, traffic monitoring radar, and so on. The effectiveness of vehicle-road collaboration significantly hinges on the RPE's perception of its surroundings. Furthermore, different types of RPEs have varying transmission requirements, ranging from several megabytes to tens or even hundreds of megabytes.

\end{itemize}

A vehicle has the capability to not only exchange processed data with neighboring vehicles but also receive processed data collected by RPE. It also can transmit sensory data, either processed or unprocessed, to the edge server located at the base station.  As necessary, a vehicle is capable of establishing communication with the edge server, e.g., in the context of updating high-precision map data. 

A RPE typically has an installation height of four to six meters, giving it a wide field of vision and perception. This elevated placement increases the likelihood of a Line-of-Sight (LoS) link between the RPE and its nearby base station. Consequently, mmWave communication with a high data rate might be utilized on this link to facilitate the transmission of a substantial volume of sensory data.

\subsubsection{\textbf{Edge Layer}}
The edge server deployed at the base station can communicate with multiple RPEs and vehicles. It may obtain either sensory data or knowledge for learning. While handling the substantial data generated by multiple devices, the edge server can also extract more valuable knowledge than devices themselves. Equipped with robust computational capability, it can produce reliable and accurate learning results, such as target recognition, classification, tracking, and so on.

V2X network is expected to effectively guarantee the communication data rate, latency and other Quality-of-Service (QoS) requirements needed for edge computing. Thus, when the computing resources of a single vehicle is insufficient, the edge server can process the sensory data collected from vehicles in order to assist the vehicles in making various decisions in time. For example, edge servers can maintain a repository of pre-trained machine learning models as vehicles do. These models are trained on extensive datasets and are proficient at executing specific tasks with high efficiency. When a vehicle encounters a new situation or scenario, it can promptly transmit relevant data, i.e., knowledge, to the edge server via a V2X links. Subsequently, the edge server can conduct real-time fine-tuning of the pre-trained model using this specific data, thereby adjusting the model parameters. This approach is especially advantageous in situations requiring adaptability, such as the changes of road condition or the shift of traffic pattern. It effectively combines the advantages of existing knowledge with the ability to adapt to current circumstances.


\subsubsection{\textbf{Cloud Layer}}

A cloud server can cater to multiple edge servers and a vast number of end devices spread across a wide geographical area. It offers various functionalities, such as training a global ML model, equipment management, system maintenance, and so on.

\subsection{Multi-Stream Driven Learning for VDR}

In VDR systems, multiple devices continually generate substantial volumes of structured and unstructured data. These devices are expected to process this data in real time and engage in appropriate collaborative actions. However, owing to constrained communication capacity, the V2X network cannot fully support real-time interaction involving the massive data from all devices. Consequently, raw data needs to be locally processed within the devices to generate the knowledge with a reduced volume before transmission. Knowledge is typically regarded of as a kind of information obtained by specific processing or learning from raw sensory data. This approach effectively reduces the number of bits that must be transmitted between devices without compromising the learning performance~\cite{Wu2019}.

Devices and servers located in different layers exhibit varying computational capabilities. Additionally, ML models differ significantly in terms of their optimization objectives and complexity, requiring distinct learning methods such as distributed and centralized approaches. This diversity results in a heterogeneous learning environment within a VDR system. Therefore, it is advisable to interact with knowledge during ML model training to efficiently leverage all computing resources using the V2X network. Once the ML model has converged, a different data stream, referred to as the \textit{model stream}, is employed to distribute the model from cloud/edge servers to devices or for local sharing among devices. Consequently, a VDR system incorporates three distinct streams to facilitate collaborative learning, i.e.,

\begin{itemize}
\item \textbf{Data stream:} This stream primarily comprises raw sensory data, including video, Lidar, and sensor data. While rich in detailed information, it tends to contain redundancy when used for learning.

\item \textbf{Knowledge stream:}  The knowledge extracted from locally mined raw sensory data is conveyed through this stream. The knowledge stream relies on V2X communication for interactive exchange between devices and servers. By utilizing this stream, demands on communication and storage resources in a VDR system can be effectively reduced while maintaining learning performance. 

\item \textbf{Model stream:} This stream contains data related to models, such as parameters for deep learning and reinforcement learning. In contrast to the knowledge stream, which focuses on sharing data generated during intermediate processes, the model stream aims to provide converged ML models capable of fulfilling final tasks, such as neural network models for video processing and reinforcement learning models for autonomous driving.

\end{itemize}

\section{V2X-Aided Collaborative Perception}

V2X communication enables the reliable and efficient exchange of information among VDR devices, laying the groundwork for intelligent collaboration. The computational capabilities across different layers of the VDR system are leveraged to offer a variety of emerging IoV services, provided the communication resources of the V2X network are fully utilized to support diverse kinds of streams. Consequently, we begin by outlining the various perception purposes and then delve into how to achieve them with the aid of V2X communication.

\subsection{Type of Perception Purposes}
Various IoV services have different requirements for environmental information, leading to distinct demands on perception. Thus,  Table~\ref{Purpose} gives several typical perception purposes in IoV~\cite{Joel2020}. 

\newcounter{mycnta}
\newcommand{\mycnta}{\stepcounter{mycnta}(\alph{mycnta})}
\begin{table}[!t]
\centering
\caption{Type of Perception Purposes}
\label{Purpose}
\begin{tblr}{
    width = \linewidth,
    colspec = {X[1,c,m]X[4.2,l,m]X[7,l,m]},
    row{1} = {font=\bfseries},
    cell{1}{2-3} = {halign = c},
    hlines,
    hline{2} = {1}{-}{},
    hline{2} = {2}{-}{},
    vline{2-3},
    column{3} = {rightsep=0pt},
}
    & Type of Purpose & Description \\
    \mycnta & Object Detection & Find all objects of a specific class in an image. \\
    \mycnta & Semantic Segmentation and Object Tracking & Assign a set of pre-defined class labels to each pixel in the image and estimate the state of multiple objects over time.\\
    \mycnta & Traffic Environment Detection & Include traffic sign detection, road/lane detection, and so on.\\
    \mycnta & 3D Reconstruction & Infer 3D geometry from a set of 2D images by inverting the image formation process using appropriate prior assumptions.\\
    \mycnta & Long-Term Autonomy & Six levels of driving autonomy and the highest level is to operate continuously and adjust automatically to road conditions. 
\end{tblr}
\end{table}


\subsection{Multi-Device Collaborative Perception}

Collaborative perception among devices, i.e., vehicles and RPEs, offers a viable solution to enhance perception performance by breaking the bottleneck of individual device perception. 

Perception requirements can vary depending on the specific purpose and may demand varying degrees of support from V2X communication. With the increase of their own computing capabilities, the vehicles can often achieve perception purposes of (a), (b), and (c) by utilizing their own sensors and processing data locally. However, in complex road conditions or adverse weather, a single vehicle may struggle to accurately perceive its surroundings. Therefore, a multi-device collaborative perception approach becomes essential to address this issue, requiring highly real-time and reliable data transmission between vehicles~\cite{Yang2020}. For perception purposes (d) and (e), vehicles can opportunistically communicate with RPE to enhance the construction of high-definition maps, known as "Mobility as a sensing service"~\cite{Chandrasekaran2018}. Additionally, the real-time constraints for such sensory data are not very stringent. It can be periodically reported to the base station through the uplink of V2I communication or transmitted with low-priority when needed.

Usually the main purposes of RPEs are to achieve (d) and (e) so as to effectively help their nearby vehicles to better achieve intelligent driving, route planning and traffic diversion, and so on. RPEs typically possess more computing and storage resources than individual vehicles, allowing them to handle tasks with high computational complexity and then distribute the results to nearby vehicles via V2X links. This approach enables vehicles to reduce their computing resource requirements for perception and allocate more of their limited resources to make real-time driving decisions. Additionally, given their cross-regional distribution, RPEs can construct wide-area, high-definition maps, facilitating more efficient route planning for vehicles. In this case, they need to engage with edge servers for information exchange, often involving high data volume transmission, e.g., through high-speed mmWave links.


\section{V2X-Aided Stream Sharing for Collaborative Learning}

\begin{figure*}[!t]
	\centering
\includegraphics[width=0.75\linewidth]{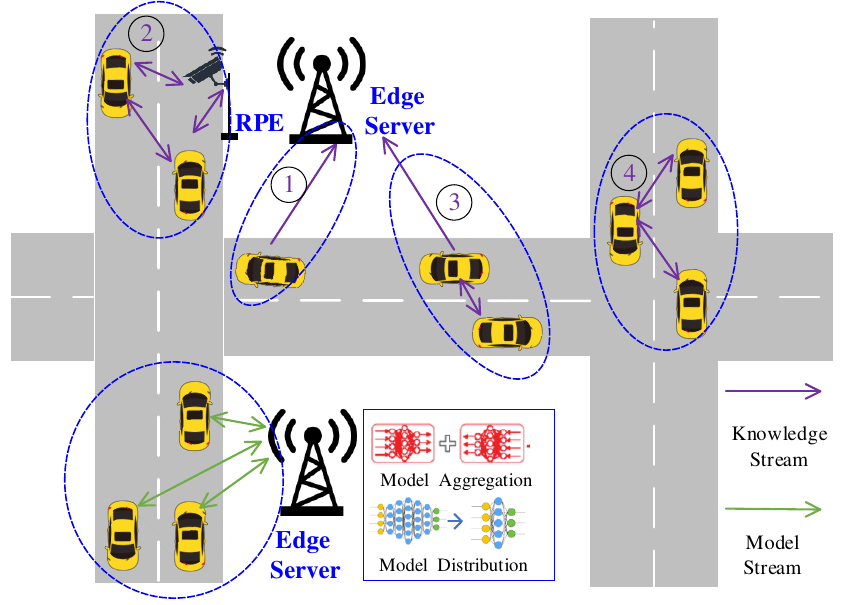}
 \caption{Illustration of stream sharing for collaborative learning in VDR.}
     \label{fig_stream_sharing}
 \end{figure*}

Due to the mobility of vehicles, the massive sensory data generated exhibits high spatial distribution. Simultaneously, various devices and servers with heterogeneous computing capabilities are scattered across the VDR system. Thus, it is crucial to effectively harness diverse computing resources for processing the massive data. Hence, as shown in Fig.~\ref{fig_stream_sharing}, we propose the strategy of V2X-aided stream sharing for collaborative learning, which depends significantly on the shared contents of knowledge and model. More details are described as follows.

\subsection{Knowledge Stream Sharing}

With the rise of ICVs, vehicles have acquired the capability to undertake a wide range of computing tasks. This allows us to leverage their abilities for tasks that require collaborative learning. For instance, when dealing with scattered training samples of unstructured data, it becomes essential for each vehicle to train a sub-model based on its local data and subsequently share this knowledge with other vehicles and edge servers. In such scenarios, knowledge is needed to be exchanged among vehicles through V2X communication.  It may include, but is not limited to, the following types, i.e.,
 
\begin{enumerate}
\item Highly informative (or highly weighted) samples for training: In the training process, certain samples are identified as crucial for learning. Sharing these samples allows devices to quickly reach global data, enabling each local model to converge collectively and rapidly to the optimal model.

\item Information through local learning: 
Devices may learn some knowledge for vehicle safety and traffic efficiency from sensory data using local learning models. This information can be shared among devices and edge servers through V2X links, helping vehicles make reliable driving decisions or facilitating efficient route planning.
 
\item Features related to collaborative learning: It can usually be shared in two scenarios. The first is between devices in the end layer, allowing them to swiftly share knowledge with one another. The second is from the end layer to the edge layer, expediting the efficient fine-tuning and inference of the model on the edge server.

\item Driving Intent: Two primary categories of driving intents exist. The former pertains to the requisites associated with the current driving task, such as the destination and anticipated travel duration. The latter encompasses common driving behaviors, including lane changes and braking. Understanding driving intents can effectively guide the collaborative driving behaviors of vehicles.
 
\end{enumerate}

The knowledge obtained can be shared via V2X networks, consequently enhancing the learning performance of ML models. Compared to traditional approaches, this strategy offers several advantages, including faster learning speed, reduced computational complexity, and lower communication overhead~\cite{NguyenCong2022}.

To facilitate knowledge stream sharing, the VDR system can employ different transmission methods according to specific QoS requirements. For knowledge with strict real-time requirements, the transmission with short packets are utilized to ensure high reliability and strict adherence to real-time constraints. Conversely, when dealing with knowledge that has loose real-time requirements, transmitting larger volumes of information can be more beneficial for finalizing model training and ensuring the completeness and diversity of knowledge.

\subsection{Model Stream Sharing}

Models within the VDR system can be shared through the model stream interaction, which occurs in two ways. Firstly, the parameters of various low-level sub-models are conveyed to higher-level entities (e.g., from devices to edge servers, and from edge servers to cloud servers) to enhance the global model. Secondly, higher-level entities also distributes performance-optimized global models to lower-level entities through the model stream~\cite{Park2021}.

The VDR system is capable of obtaining a global ML model through knowledge stream sharing on V2X links. However, when dealing with complex models for intelligent driving, there is often a challenge in efficiently aggregating and distributing these models.

In the context of model aggregation, different approaches can be employed based on the nature of the learning tasks and their specific optimization objectives. Model averaging is a highly effective technique for learning tasks characterized by convex optimization objectives. However, for non-convex problems such as deep learning, integration learning is the preferred approach due to its advantages. This involves solving the same problem using multiple models, often referred to as weak learners, and then combining their outputs to achieve improved results.

When it comes to model distribution, the key goal is to minimize communication overhead while maintaining optimal learning efficiency. As a result, it is essential to make model distribution as latency-minimal and data-efficient as possible. One approach to reduce the size of the transmitted model is by employing the low-rank decomposition of model matrix. Furthermore, the model can be further simplified through low-precision quantization or the random discarding of model parameters.

\section{V2X-Aided Secured Collaboration}


\begin{figure*}[!t]
	\centering
\includegraphics[width=0.8\linewidth]{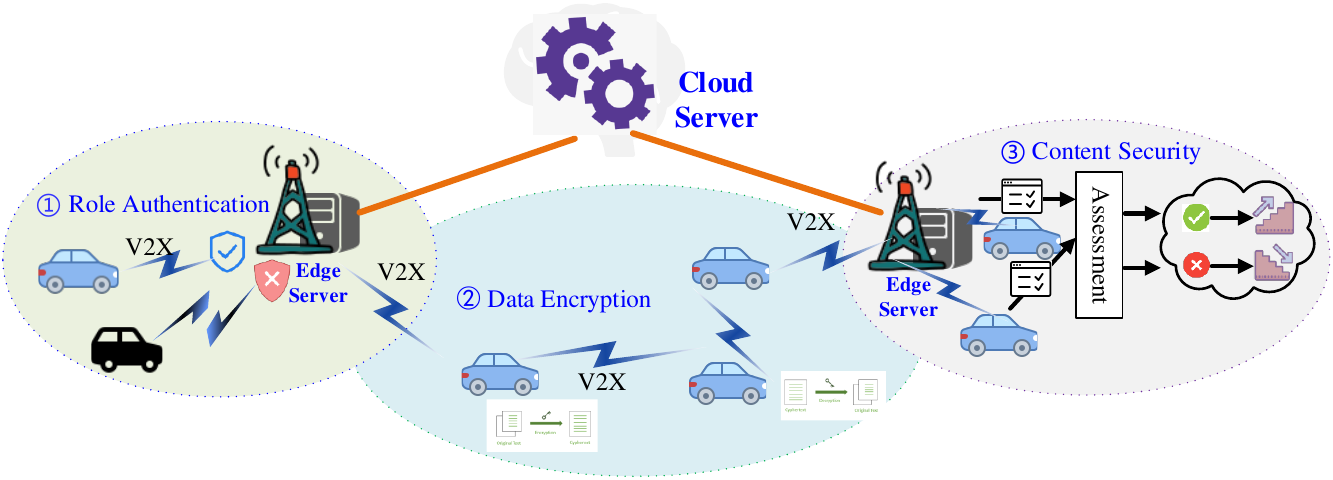}
 \caption{Illustration of Secure Collaboration for VDR.}
     \label{fig_secure_communication}
 \end{figure*}

 In order to realize the collaborative perception and learning mentioned above, it is essential to guarantee the safe interaction of various streams with the help of V2X communication in the VDR system. To achieve this goal, we have identified several crucial aspects that demand secure collaboration in the VDR system, as depicted in Fig.~\ref{fig_secure_communication}.

\subsection{Role Authentication}

Collaborative participants anticipate efficient and reliable driving services in the VDR system. Nevertheless, the presence of malicious actors might significantly undermine the quality of service. For instance, when these malicious participants fraudulently join collaborative perception by assuming false identities, they can interfere with data and knowledge streams using tactics like replay attacks or Denial-of-Service attacks against the server.

Typically, servers located in the edge and cloud layers are actively engaged in collaborative perception and learning with honesty and integrity. In the end layer, while RPEs may occasionally encounter issues like missing sensory data, they generally contribute their computing resources transparently and participate in various stream interactions, thus establishing themselves as secure devices. However, a large number of vehicles pose a challenge, as attackers can potentially impersonate their identities to join collaborative activities, introducing concerns related to legitimacy and role security. To guarantee a secure and reliable participation in collaborative perception and learning, it is imperative to implement a robust identity authentication mechanism within the VDR system~\cite{iovrole}. 


When a vehicle initiates a request to participate in collaboration, it can be authenticated based on both static and dynamic data,  which encompass vehicle registration records and real-time environmental conditions, respectively. This verification process is typically carried out by the edge server in proximity to the vehicle, involving the use of identity data to confirm the legitimacy of the participant. To ensure the integrity and credibility of this authentication process, blockchain technology can be employed to securely store and manage these data. By deploying blockchain nodes across multiple edge servers and distributing authentication functions among them, the risk of malicious participants infiltrating the collaboration can be effectively mitigated.

\subsection{Data Encryption}

The VDR system relies on V2X networks to facilitate the interaction of various streams among devices and servers. However, the mobility of vehicles makes V2X networks susceptible to various attacks such as data replay, tampering, and forgery attacks~\cite{iovcom}. Thus, ensuring the security of different streams during their transmission in V2X networks is crucial to enable effective knowledge sharing and collaborative learning in the VDR system.

Traditionally, data is encrypted before transmission to enhance security. However, this encryption process introduces additional computational and transmission overhead, which can negatively impact the efficiency of collaborative perception and learning. Consequently, it is essential to choose appropriate encryption methods based on the characteristics of different kinds of streams, effectively reducing overhead and enhancing collaboration efficiency. For instance, for knowledge and model streams with relatively small data volumes, complex encryption methods with high security and redundancy can be employed to protect wireless transmission from malicious attacks. In contrast, data streams with larger data volumes can benefit from simpler encryption methods with lower overhead to prevent V2X network congestion.

\subsection{Content Security}







After the participant receives the streams through V2X networks, it is also necessary to check whether the data content itself in the stream is reliable. In the VDR system, conducting security assessment of the content based on the accumulated reputation is a feasible method to provide a reliable interaction environment among vehicles~\cite{yangzhepaper}. The reputation assessment is frequently indirect and needs to be obtained based on quantitative analysis of the reported contents of the participants, and the positive contribution metrics need to be customized by the scenario characteristics and learning objectives. Meanwhile, the participant reputation obtained by quantitative analysis has to be stored in a traceable, distributed, and secure manner, thus guaranteeing efficient collaboration. 

\begin{figure*}[!t]
	\centering
\includegraphics[width=0.9\linewidth]{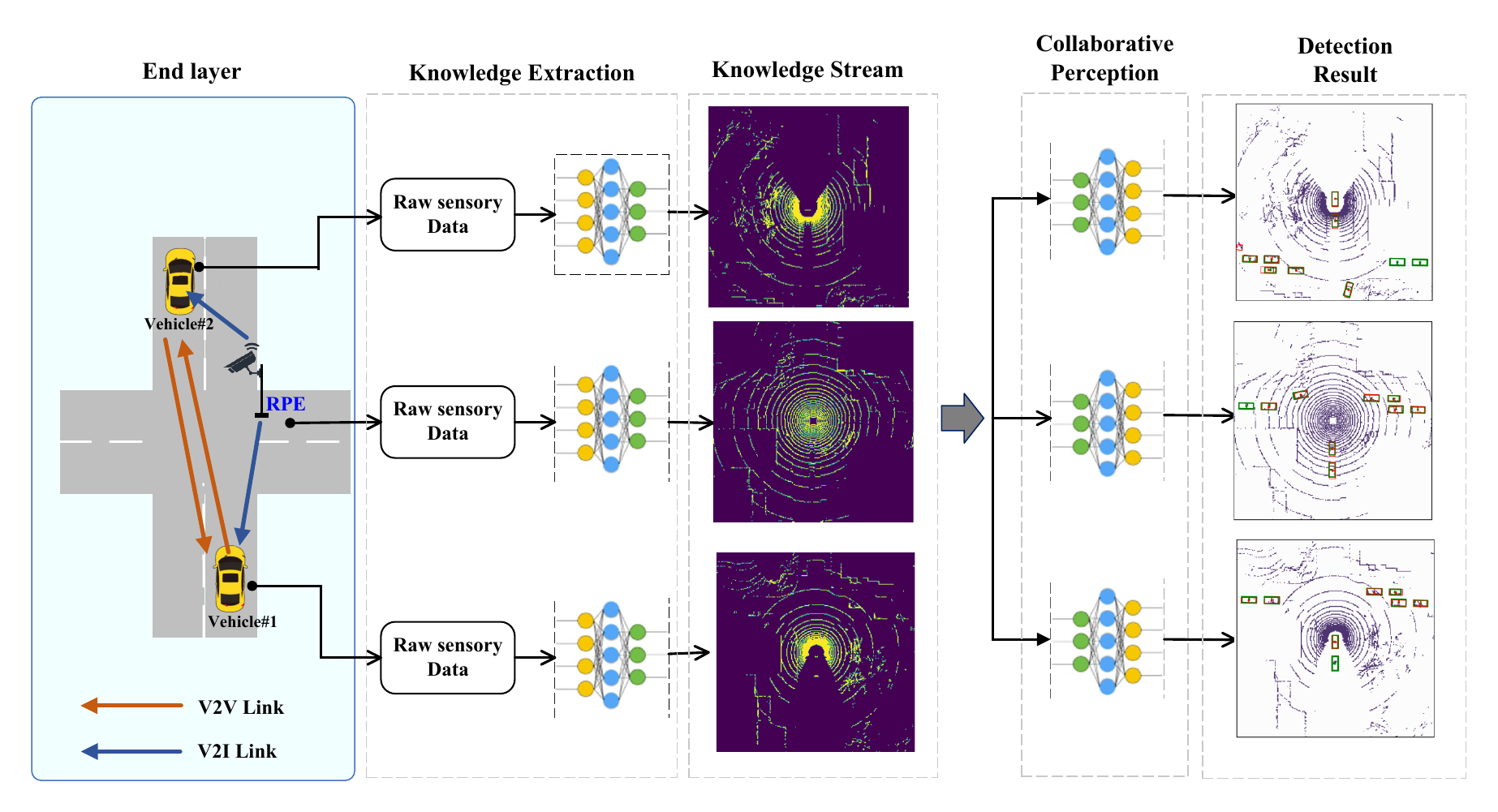}
 \caption{Illustration of a VDR-based collaborative perception scheme.}
     \label{fig_case_study}
 \end{figure*}

In general, a variety of techniques are used to assess participant reputation in the VDR system in accordance with specific characteristics of the streams, i.e.,
\begin{itemize}



\item \textbf{Data Stream Interaction}: In the case of data streams with large volumes and potential redundancy, assessing the security of the data is primarily handled by the edge server. It does so by accumulating long-term reputation values to evaluate the trustworthiness of the perception content.

\item \textbf{Knowledge Stream Interaction}: Knowledge streams typically have smaller data volumes but contain critical information. During collaboration, multiple vehicles and RPEs cross-verify each other's knowledge. RPEs or edge servers can collect these verification results and make comprehensive judgments regarding the authenticity of the knowledge reported by participants. These results are then broadcast on the V2X network, serving as reference reputation for different participants.

\item \textbf{Model Stream Interaction}: Participants employ a portion of their local data as a verification dataset to validate the performance of the received model stream, including aspects like model accuracy and loss. Based on these verification results, their assessment of whether the received models are secure and meet performance benchmarks contributes to determining participant reputation.

\end{itemize}



\section{Case Study for Collaborative Perception}

To illustrate the effectiveness of VDR systems, we have implemented a VDR-based collaborative perception scheme aimed at enhancing object detection performance in a simulation platform. For this study, we consider an urban intersection scenario in which a Reference Point Entity (RPE) is located at the center. Meanwhile, multiple vehicles, each following its driving route, are concurrently present at this intersection. To conduct our experiments, we utilize an extensive perception dataset, i.e., V2X-Sim, generated by SUMO and Carla~\cite{V2X-Sim}. The simulations are carried out on an X86 PC station equipped with 2 Intel Xeon Silver 4210R @2.40 GHz, 20 core CPUs, 64 GB RAM, and 1 NVIDIA RTX3090 GPU. As illustrated in Fig~\ref{fig_case_study}, the knowledge sharing among devices including vehicles and RPE happens at the edge layer via V2X communications. The knowledge on each device is extracted from raw sensory data and transformed to Bird’s-Eye-View (BEV) representations.

The mean average precision (mAP) metric is selected as the measurement for perception performance. Additionally, Intersection-over-Union (IoU) thresholds of 0.5 and 0.7 are adopted in our simulations. As shown in Table~\ref{Result}, it is evident that VDR-based collaborative perception outperforms the single-vehicle perception scheme in cases of 1 RPE and 2 vehicles, resulting in a substantial enhancement. Specifically, there is a 9.7\% improvement for mAP@IoU=0.5 and a 10.2\% improvement for mAP@IoU=0.7, respectively.

\begin{table}[!t]
\centering
\caption{ Comparison of Perception Performance under Urban Intersection Scenario (1 RPE and 2 Vehicles)}
\label{Result}
\begin{tblr}{
    width = \linewidth,
    colspec = {X[7,l,m]X[3.5,c,m]X[3.5,c,m]},
    row{1} = {font=\bfseries},
    cell{1}{2-3} = {halign = c},
    hlines,
    hline{2} = {1}{-}{},
    hline{2} = {2}{-}{},
    vline{2-3},
    column{3} = {rightsep=0pt},
}
    & mAP@IoU=0.5 & mAP@IoU=0.7 \\
    Single-vehicle scheme & 46.08\% & 38.69\%  \\
    VDR-based collaborative scheme & 50.56\% $\;\;\;\;\;\;\;\;\;$($\uparrow$ 9.72\%)  & 42.67\% $\;\;\;\;\;\;\;\;\;$($\uparrow$ 10.29\%)
\end{tblr}
\end{table}

\section{Conclusion and Outlooks}
\label{sec:Conclusion}

This paper has proposed a collaborative learning system based on multiple kinds of streams to efficiently utilize the communication and computing resources of multi-layer vehicular networks. Three different kinds of streams, each with its own characteristics and interacting via V2X networks, can effectively support a variety of learning tasks. Having presenting the framework of the VDR system, we further study  how collaborative perception and learning can be used to implement various IoV applications. Moreover, a few new schemes for role security, transmission security, and content security have been discussed in order to ensure secured collaboration in the VDR system. Once the aforementioned issues have been satisfactorily resolved, the VDR system can efficiently support a variety of new IoV services. Our case study has demonstrated that knowledge sharing in the VDR system can significantly improved perception performance.

Building upon the VDR framework outlined in this paper, our next steps will center on the design and implementation of specific optimization algorithms for collaborative driving. Firstly, we will establish a method to evaluate the Value of Knowledge (VoK), enabling a quantitative measure for knowledge stream interaction and facilitating the implementation of diverse collaborative strategies. Then, we will study collaborative perception schemes based on deep reinforcement learning (DRL), aimed at maximizing the overall VoK to achieve profound environmental perception. Additionally, our research will place significant emphasis on the collaboration between edge and devices based on knowledge, allowing for the share of machine learning models that support collaborative driving. Furthermore, collaborative security will be enhanced by using technologies such as blockchain in our upcoming work. We are also committed to conducting an in-depth analysis of the trade-off between security measures and computational complexity of new collaborative security schemes.

%
\bibliographystyle{IEEEtran}
\bibliography{IEEEabrv,Ref}
%
%

\end{document}